# PLANS FOR THE SPALLATION NEUTRON SOURCE INTEGRATED CONTROL SYSTEM NETWORK[*]

W. R. DeVan, E. L Williams Jr., ORNL, Oak Ridge, TN 37830, USA

Abstract

The SNS control system communication network will take advantage of new, commercially available network switch features to increase reliability and reduce cost. A standard structured cable system will be installed. The decreasing cost of network switches will enable SNS to push the edge switches out near the networked devices and to run high-speed fiber communications all the way to the edge switches. Virtual Local Area Network (VLAN) technology will be used to group network devices into logical subnets while minimizing the number of switches required. Commercially available single-board computers with TCP/IP interfaces will be used to provide terminal service plus remote power reboot service.

## 1 INTRODUCTION

The Spallation Neutron Source (SNS) is an accelerator-based neutron source that will be used primarily for neutron scattering R&D. National laboratories participating in the project and their areas of responsibility are:
- LBNL: ion source and front end
- LANL: linac
- TJNAF: cryogenic helium liquifier (CHL) and linac cryomodules
- BNL: accumulator ring and beamlines
- ORNL: target and conventional facilities
- ANL: neutron instruments

In general, these labs have responsibility for the control systems corresponding to their technical system. ORNL has responsibility for implementing the control system communications network that ties these control systems together.

SNS will use the Experimental Physics and Industrial Control System (EPICS) for accelerator, target, CHL, and conventional facilities controls. This distributed control system for SNS has been dubbed the "Integrated Control System" (ICS). An integral and critical part of the ICS is a large-scale local area network dedicated to control system communications. This paper describes plans for the ICS communications network.

## 2 NETWORK OVERVIEW

The SNS control systems group will maintain four physical Ethernet networks. The primary network supports EPICS channel access and PLC communications. The remaining three networks support maintenance of the primary network. One of these provides "out of band" maintenance access to communication room switches. (Maintenance access to edge switches is handled "in band" as described later). There are also two networks to support "sniffing" of channels on the primary network.

A conventional structured cabling system is used. The main hub will be in the Front End (FE) Building (the first building constructed). Fiber optic "backbone" cables will radiate out from the hub to communication rooms (14 total) scattered throughout the facility. These cables will provide fibers for other SNS communication systems as well (e.g. the timing system and the administrative network).

A description of the primary network follows. A conventional hierarchical switch architecture is used. Gigabit Ethernet (1000BASE-SX and -LX) is used for backbone communications. Redundant core switches in the FE Building link to a layer of aggregator switches in communication rooms. Gigabit Ethernet is then extended from the aggregator switches to a third layer of access switches in the service buildings. These switches provide 10/100BASE-TX service to IOCs, PLCs, OPIs, and a fourth layer of switches. The fourth layer of switches services beam diagnostic devices and temporary connections. The network is fully switched; no repeaters are used.

Table 1 provides a summary of hosts and network drops connected to the primary network.

In the interest of minimizing the number of network switches, separation of subnets will be maintained by the use of 802.1q VLAN technology (vs. each subnet having its own set of switches). The primary control system network includes separate VLANs for accelerator, target, cryogenics, and conventional facilities controls. There is also a corresponding set of maintenance VLANs for accessing edge switch console ports and IOC console ports. (See more below).

---

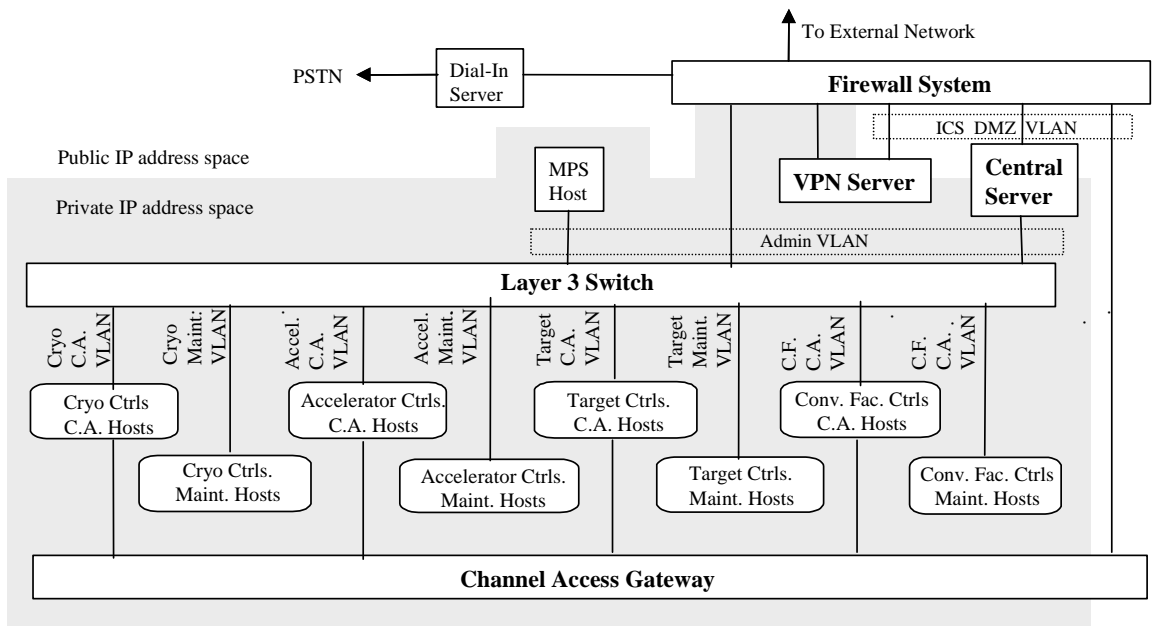

Figure 1: ICS Control System Network

Table 1: ICS Network Hosts and Ports

| Host Type | Quantity |
|---|---|
| Operator Interfaces (OPIs) and Servers | 38 |
| I/O Controllers (IOCs) | 178 |
| Programmable Logic Controllers (PLCs) | 118 |
| Beam Diagnostic Devices | 558 |
| Misc. (Scopes, function generators, power monitors, etc.) | 36 |
| Terminal Server / Remote Power Rebooters | 200 |
| Utility network drops at IOCs and PLCs | 180 |
| Tunnel network drops | 117 |
| Total no. of ports | 1425 |

To enhance reliability, the cryogenic control system subnet is physically separate from the other VLANs. This subnet will be able to continue operating should any non-cryo communications equipment fail.

The primary network's core switches are Cisco model 6509. Second and third layer switches are Cisco 3500 series. Fourth layer switches are typically Cisco 2950 series.

It should be noted that the ICS network performs no personnel safety functions. Nor is it directly responsible for machine protection. It is however utilized in the configuration of the machine protection system, and does provide some "defense in depth" for machine protection. Consequently adequate security must be provided to prevent unauthorized persons from interfering with these functions.

## 3 NETWORK MAINTENANCE

Remote terminal service is provided to network switch console ports in order to reduce "mean time to repair" of network problems. This service is implemented via Rabbit Semiconductor single board computers (at SNS commonly referred to as "Rabbits"). These boards feature a 10BASE-T Ethernet port, 4 digital inputs, 4 digital outputs, RS-232 port, and 512K flash memory. (These boards will also provide terminal and remote power reboot service to IOCs). A software library, including a telnet routine, is provided with the boards. The ORNL Hollifield Radioactive Ion Beam Facility (HRIBF) has modified the telnet routine to include a password-protected control sequence to initiate remote power reboot of the connected device. Remote power reboot will not be used for network switches until it is determined that the rebooter is more reliable than the switch.

An out-of-band network supports terminal service to communication room switch console ports. This out-of-band network has a "star" topology with a core switch in the FE Building. Backbone fibers provide connections from the communication rooms to the FE Building.

Terminal service to edge switches is handled "in-band". The RS-232 console port of each edge switch is connected to a Rabbit device. The Rabbit's ethernet port is connected to another nearby switch. Maintenance personnel can connect to the console port of the edge switch via a telnet connection to the Rabbit.

Rabbits will be connected to dedicated maintenance VLANs.

Two sniffer networks are provided for remote sniffing of switch channels. SNS has dedicated one port on each switch for port mirroring. For the first sniffer network, mirror ports on aggregator switches in communication rooms are connected directly to the FE Building communication room via backbone fibers. A gigabit Ethernet sniffer in the FE Building can then monitor any primary network gigabit Ethernet channel. For the second sniffer network, a similar architecture is used for sniffing 10/100BASE-TX channels. However, there are too many mirror channels in this case to directly connect every one to the FE Building communications room. Instead, each mirror channel is connected to the nearest communication room (via a 100BASE-TX link). Then one fast Ethernet connection is provided from each communication room to the FE Building. Maintenance personnel will need to manually patch in the channel they want to sniff.

The use of commercial network management software is planned but the software has not been selected yet.

## 4 ADMINISTRATIVE SERVICES

A design goal is for the control system to be able to operate SNS independently of any temporary loss of resources outside the ICS network perimeter. Consequently any administrative services required for network operations are included as part of the control system network.

Current plans are for redundant DNS and DHCP servers to provide name service and address management. A RADIUS server will provide authentication for the dial-in server and VPN access. LDAP will be used for account management.

Central servers will be backed up over a dedicated back-up VLAN. This requires all central servers to have at least two Ethernet interfaces, one for normal network traffic and another for back-up VLAN access.

## 5 NETWORK SECURITY

The ICS network will utilize a private IP addressing scheme per RFC 1918. A Class B IP address space will be provided. Hosts within the private address space will not be announced outside the ICS network perimeter.

A firewall will be provided to restrict access to the ICS network. Restrictions will be implemented via standard commercially-available firewall components (e.g. packet filtering, proxy service, and/or stateful inspection). Access control lists (ACLs) on the "Layer 3" core switches will be used to control access between VLANs.

EPICS offers a standard set of security features, including the ability to designate which hosts and/or users can access a given control parameter. To help prevent unauthorized access to control functions, SNS plans to limit direct channel access across VLANs. An EPICS "channel access (CA) gateway" will provide indirect access to control parameters across VLANs and to the outside world.

Remote access to the ICS network will be allowed but will be strictly controlled. Allowable methods of access include SSH access to central servers, VPN access, and dial-in service. SSH and VPN encrypt network traffic outside the ICS network perimeter, so passwords remain protected. Dial-in access will be password protected, monitored, and logged.

Due to the rapidly changing product offerings in the network security arena, specification of network security equipment is being put off as long as possible. We expect to have these devices specified and prototyped by Spring of 2002.

## 6 NETWORK RELIABILITY

The primary network's core switches are redundant. Links from the core switches to aggregator switches in communication rooms are redundant as well. The aggregator switches and all switches below that layer are not redundant.

For core switches and aggregator switches, two independent power circuits (one being UPS) will feed each switch. Edge switches will be fed from a single power circuit (UPS where available). Air conditioning to communication rooms is not redundant, so network switch temperatures will be monitored via a network management workstation in the main control room.

## 7 PLANS AND SCHEDULE

A prototype network has been implemented at the temporary SNS office building, with required features being added in an evolutionary fashion. It is planned for all required features to be prototyped at this office location before moving the network to the site.

Construction of SNS buildings has started and will continue through 2004. The first segment of the control system network will become operational in July 2002. Network equipment will be added as buildings become available. Completion of the ICS network is expected by the end of 2004.